*The recent Italian regulations about the open-access availability of publicly funded research publications, and the documentation landscape in astrophysics*

*Monica Marra* [1]

[1] *INAF (Istituto Nazionale di Astrofisica), Osservatorio Astronomico di Bologna, via Ranzani 1, I – 40127 Bologna, Italy;*
email: monica.marra@oabo.inaf.it

**Abstract**. In October 2013 Italy enacted Law n.112/2013, containing the first national regulations about the open-access availability of publicly-funded research results (publications). The impact of these new regulations with the specific situation of that open-access pioneering discipline which is astrophysics, has been considered. Under a strictly technical point of view, in the light of the new dispositions the open nature of a part of the astrophysical scholarly literature which has been made available online free to the reader during the last twenty years, might be questionable. In astrophysics, most of the journal articles are published by a very small number of scientific societies and organizations. The copyright policies of these major publishing bodies have been collected and analyzed, with regard to the main requirements of the Italian law about open access. Most of these policies are sufficiently liberal for an entirely compliant open access to be provided and scientists would benefit from knowing these details more extensively. Some possible ways to make astrophysicists' scholarly dissemination entirely compliant with the Italian law requirements are considered.

On October 7th, 2013 Italy achieved an important attainment enacting Law n.112/2013, which is the first national law concerning open access.

The law – which in fact pertains to a number of issues related to cultural heritage - had initially been approved in a different formulation (Decreto-Legge n.91 dated August 8th, 2013). During the passage through Parliament, the first stage of the action was subjected to changes, which according to authoritative commentators (De Robbio 2013; Caso 2013) partly deflected the text about open access from the European Union Recommendation on the subject, issued in July 2012.

Law 112, under article 4, decrees:

- OA is mandatory for research outputs/publications derived from research projects which have been financed by at least 50% with public funds;

- if the publication venue publishes at least 2 issues/year (i.e., according to Caso 2013: books are excluded).

- It may be provided a) either on the publisher's site or b) on an institutional or disciplinary repository, by depositing the publication's accepted version within 18 months (STM area) or within 24 months (SSH area) from publication.

- It must be provided at no additional cost.

During the period that shortly preceeded and followed the enactment of Law 112/2013, a series of important initiatives seemed to turn the Italian experts' and concerned people's attention back from the new regulations to the European Recommendation. These initiatives included a) the commitment, signed by the then-Minister for University and Research Massimo Bray, aimed at harmonizing the new law with the European Recommendation (October 3rd, 2013) and consequent actions (i.e., the programme Scientific Independence of young Researchers (SIR) 2014, with open access mandatory clause in the same terms as in Horizon 2020, mentioned below); b) the approval of the 70-billion-euros research financing programme Horizon 2020 at the European Commission, in Autumn-Winter 2013, with European-type open-access clauses; c) the publication of some relevant national statements and recommendations, with a leading role carried out by national academic coordination bodies, also supported by the Presidents of some important national research Institutions, including INAF [ 1 ]. These recommendations and statements seem to lean towards the green road to open access and encourage academic and research institutions to issue OA policies and mandates for their research personnel.

In a general perspective, it may be useful to recall that astrophysics is peculiar among other disciplines, with regard to its approach to the dissemination of research outputs. Astrophysics has actually been so deeply crossed by "the preprint culture" during the last twenty years, that in the opinion of most scientists at international level an open-access policy is considered to be already practised (Harley 2010), the proof being generally indicated in the very large adoption of the practice of posting papers on the ArXiv server in a pre-publication stage.

The literature is not unanimous about the percentage of published astrophysical papers already available in open access mode - which is probably related to the different self-archiving practices adopted by researchers according, typically, to different astrophysical subfields. Actually, archiving of the preprint version of papers occurs as well as archiving of the accepted version does; the (limited) posting of post-print versions as well makes this lively scenario even more rich and motley.

Anyway, although it seems unquestionable that astrophysicists' perception of a widely open-mode available disciplinary literature is well-grounded, we might profitably ask ourselves whether researchers on one side, and regulations about open access on the other, are actually considering the very same object.

In fact the Italian law - as well as other national laws concerning open access (Caso 2013) - considers texts as endpoints of the respective scholarly pathways rather than possible stages of research elaboration and dissemination. Thus, regulations require making free to the reader *the papers' final accepted versions* within a given period - either by publishers on their website, or by authors through systematic deposit on a repository. As unexpected as it may seem, a significant fraction of the impressively long "green OA" road which has spontaneously been taken by astrophysicists during the last twenty years might thus result to be non-compliant with the new law requirements, if considered from a strictly technical point of view.

One of the foremost questions for those technically concerned with open access in general is the following: for open access to be provided on authors' side, do publishers let authors manage their manuscripts freely enough? In particular, and key to our perspective: what do publishers allow authors to do with the final accepted versions of their papers?

The crux is authors' rights as stipulated in copyright agreements with publishers, therefore it is essential to be aware of the main journals' policies with regard to this issue.

In astrophysics, "more than 90% of the original research" is published by a very limited number of renowned international journals (Bertout 2012), almost entirely included in the first and second quartile of Thomson Reuters' Journal Impact Factor (JIF) ranking for Astronomy and astrophysics. I have analysed this set of major journals, which includes ApJ (JIF for 2013: 7/59), ApJL (10/59), ApJSS (2/59), AJ (15/59), MNRAS (11/59), A&A (13/59), PASP (17/59).

[ 1 ] On March 21st, 2013 the Conference of Rectors of the Italian Universities (CRUI), together with the Presidents of some important Italian research Institutions, signed the *Position statement on open access to the results to research outputs in Italy* (available in English at http://www.cnr.it/sitocnr/Iservizi/Biblioteche/Position_statement_OA_en.pdf , last visited Nov. 10[th], 2014). This statement is envisioned in a European perspective and commits the signers to coordinating initiatives aimed at i) promoting the creation of open repositories, that be interoperable according to international standards; ii) "encouraging researchers to make their research results (data and publications) available in OA journals or Institutional or subject repositories. Research results deposited in Open Access repositories should be made available in their post-print or publisher's version upon publication, and no longer than 12 months after their publication;" [italics mine]; iii) adopting institutional policies for making deposit mandatory. In January 2014, CRUI issued the Declaration *L'"accesso aperto" alle pubblicazioni scientifiche* (signed jointly with the Italian National University Council (CUN)). The Declaration is critical about the maximum embargo periods stated by the new law and encourages the choice of the "green OA" option (https://www.cun.it/provvedimenti/sessione/140/dichiarazione/dichiarazione-del-15-01-2014 - in Italian, last visited Nov. 10[th], 2014).

All of the journals in this set have both a publisher, and a learned society or scientific organisation which is in fact the reference point for the journals' scientific policy and, which is most important from the present point of view, owns the copyright for the journals' publications and may have a copyright policy of its own, different from the publisher's one. The learned societies or organisations involved are the American Astronomical Society (published by the Institute of Physics Publishing), Royal Astronomical Society (published by Oxford University Press), European Southern Observatory (EDP Sciences), Astronomical Society of the Pacific (University of Chicago Press).

Thanks to the precious collaboration of these learned societies and organisations, the major journals' policies have been individually checked and a comprehensive view of the options which authors are given was gained - as at Spring 2014.

About the so-called "gold road" to open access – which is in fact notoriously questioned among experts as a recommendable choice for institutions -, three versions turned out to be available.

ApJ, ApJL, ApJSS, AJ, PASP and A&A offered a 12-months delayed open access to the entire content of these serials, free to the reader. The time issue here might arise as a critical element in case regulations or mandates (not to mention the authors' wish) asked for a shorter period for the openness to be provided on the publishers' side. The article processing charge (APC) option was not provided – except for A&A -, but the AAS "welcomes discussions" with interested institutions.

The European journals practised a threefold policy. On one hand, A&A and MNRAS provided niches of their content free to the reader upon publication, while making the remaining full-texts openly available after a period which ranged from twelve months (A&A) to the impressive threshold of three years (MNRAS).

On the other hand, they provided research institutions with the APC option, which corresponds to 400 euros per article for A&A and more than five times as much for MNRAS (as at November 2013).

Obviously, an expenditure issue may arise in some of these cases.

For as much as the "green road" is concerned, ApJ, ApJL, ApJSS, AJ and PASP let authors post the accepted versions of their articles freely on repositories, "whenever they like" for the AAS and also immediately - provided there is an institutional mandate requiring deposit in fewer than twelve months - in the case of PASP.

"Astronomy and Astrophysics" let authors self-archive the accepted version of their papers before publication, but asked researchers to substitute this version with the published one, once the latter is available.

MNRAS let authors self-archive exclusively a PDF version of the final published article, upon publication (as at November 2013).

Bearing in mind all of these copyright policies and the law requirements; considering the citation advantage for papers disseminated in a pre-publication stage; and being notorious that astrophysicists' scholarly routine is in most cases already inclusive of a timely, pre-publication self-archiving of papers on a repository; we might conclude by endorsing the Green Road to open access as a first choice option, thus supporting the recommendation issued by the Italian Conference of Rectors (CRUI), which was signed by the Italian National Institute for Astrophysics on October 4th, 2013 [ 2 ].

---

[ 2 ] CRUI and some relevant Italian public research Institutions recommend " [...] the adoption of policies and institutional regulations which make deposit in their institutional repositories mandatory; in case the latter don't exist, publications and data should be deposited in other institutions' repositories or in disciplinary OA repositories".

I

In case the economics of these processes are a matter of concern, the peculiar modes of dissemination practised by astrophysicists, which have hitherto been characterized by the "productive coexistence" (Henneken et al, 2007) of established journals on one side, and very wide networks of versioned, free online availability on the other side, might preserve the sustainability of the learned societies' activity, or in some cases (MNRAS) hopefully make it more generally affordable to research institutions.

A final, additional consideration deserves attention. In spite of their central importance within astrophysics, the refereed journal articles seem in fact to represent a minority share of the discipline's research output (less than 40% according to the sample study by Marra, 2013). In the spirit of making open access progressively more common practice, the use of a Current Research Information System (CRIS) as an official tool for compliance with the OA provisions might open the way to a similar treatment for the other research and technology output types which are or might be hosted by this kind of tool, and which cooperate significantly in the shaping of the astrophysical documentation landscape.